\documentclass[aps,prl,preprint,superscriptaddress]{revtex4}

\usepackage{graphicx}
\usepackage{amssymb}

\begin{document}

\title{Exact results for Casimir forces using Surface Impedance: Nonlocal Media} 

\author{R. Esquivel}
\email[Corresponding author. Email:]{raul@fisica.unam.mx}
\affiliation{Instituto de F\'{\i}sica, Universidad Nacional Aut\'onoma
 de M\'exico, Apartado Postal 20-364, D.F. 01000,  M\'exico}

\author{C. Villarreal}
\affiliation{Instituto de F\'{\i}sica, Universidad Nacional Aut\'onoma
 de M\'exico, Apartado Postal 20-364, D.F. 01000,  M\'exico}

\author{W. L. Moch\'an}
\affiliation{ICentro de Ciencias F\'{i}sicas, Universidad Nacional Aut\'onoma
 de M\'exico, Av. Universidad S/N, Cuernavaca, Morelos 62210,,  M\'exico}

\date{\today}

\begin{abstract}
We show that exact results are obtained for the calculation of Casimir forces between arbitrary materials using the concept of surface impedances, obtaining in a trivial way the force in the limit of perfect conductors and also Lifshitz formula in the limit of semi-infinite media. As an example we present a full and rigorous calculation of the Casimir force between two metallic half-spaces described by a hydrodynamic nonlocal dielectric response. 
 \end{abstract}

%\pacs{12.20.Ds}
%\keywords{}

\maketitle

 In recent 
years  experimental studies of Casimir vacuum forces 
between metallic surfaces have reached the necessary accuracy to test in detail 
the theoretical predictions \cite{lamoreaux,mohideen1,mohideen2,chan} posed originally by Casimir in 1948 
for perfectly-conducting parallel plates \cite{casimir}. 
Indeed, the accuracy reached in the experiments has forced to consider  
detailed properties of the materials such as absorptivity, rugosity, or
finite  temperature effects \cite{bordag,kikat,sernelius}.  
The study of vacuum forces between realistic materials was pioneered
by Lifshitz in 1956, who proposed a macroscopic  theory for semi-infinite 
slabs described by a dielectric function $\epsilon(\omega)$ \cite{lifshitz}. 
Lifshitz formula, which reduces to the Casimir result
in the limiting case of perfect conductors, has been successfully employed
in a number of experimental situations. 
Different authors have elaborated
alternative derivations of Lifshitz formula that permit a simpler, 
more transparent approach to vacuum forces in realistic materials \cite{otros}. Among  several proposals to calculate the Casimir force  the impedance approach was
 employed for the first time by Mostepanenko and Trunov \cite{trunov} to derive the Lifshitz 
formula in an approximate fashion.  
In a series of investigations of vacuum forces in realistic materials, 
both at zero and finite temperature, Klimchitskaya \cite{kika} et al. have found small  discrepancies in the spatial behaviour of the forces when
calculated according to the impedance or the Lifshitz approach. In 
those papers it is argued  that the discrepancies arise from the approximate 
nature of the impedance concept. Indeed, within the particular context of the 
optics of metals, the surface impedance is usually derived by taking approximations valid only for good conductors below their plasma frequency, and 
it is interpreted in terms of induced surface currents \cite{landau}. 
However, such approximations are unnecessary and the concept of surface impedance can be straightforwardly applied to arbitrary materials \cite{stratton,halevi,exciton}. 
In an equivalent framework, Kats \cite{kats} introduced the reflection amplitude 
coefficients $r_a^\alpha$ ($\alpha = s,p$) for metallic media to derive also an 
approximate version of Lifshitz formula to study the influence of nonlocality on van der Waals interactions in a semi-quantitative way. Noticeably, he  stated
incorrectly that the reflection coefficient cannot be expressed merely in terms of the surface impedance for dielectrics.  
 Recently, a more rigorous  derivation of the Casimir force in terms of the reflection amplitudes was discussed by Reynaud and collaborators \cite{reynaud} using a $S$-matrix formalism. 
 
The surface impedance $Z$ is $defined$ as the ratio of the complex electric
and magnetic field components in the direction parallel to the surface of incidence
of an electromagnetic wave, evaluated just inside the surface \cite{stratton,halevi}.
For an $s$-polarized wave incident on a 
planar surface $z=z_0$ moving (at an angle $\theta$) towards the positive 
$z$ direction, the impedance boundary condition is 
\begin{equation} \label{impeds}
{\bf E}(z_0^-) = Z^s( {\bf H}_t(z_0^-) \times {\bf \hat{z}}),
\end{equation}
while for a $p$-polarized wave, the corresponding definition is 
\begin{equation}\label{impedp}
Z^p{\bf H}(z_0^-) =  {\bf \hat{z}}\times{\bf E}_t(z_0^-) ,
\end{equation}
where ${\bf \hat{z}}$ is a normal vector pointing inside the surface, and 
$z_0^-$ denotes a position immediately before the $z=z_0$ interface.
A main advantage of this concept is that it relates only external tangential fields, 
without the need of involving 
the internal degrees of freedom of electromagnetic field inside the material.
 Equations (\ref{impeds}) and (\ref{impedp}) are functional definitions of the surface
impedance, so that they are  $exact$ and valid, not only for perfect conductors, but also
for real metals and insulators.  Indeed, from these definitions the reflection coefficients of the electromagnetic fields are obtained as 
\begin{equation}
\label{refl}
r_a^s =\frac{Z_a^s - Z^s_0}{Z^s_a + Z^s_0} \ \ \ r_a^p =\frac{Z_0^p - Z^p_a}{Z^p_0 + Z^p_a},
\end{equation} 
with $Z^s_0 = q/ k_0$, and $Z^p_0 =  k_0/q$ and $Z^p_a = k_a c/( \epsilon_a \omega )$, $Z^s_a =\omega/k_a c$.
These definitions in terms of the impedance are exact. For a local semi-infinite medium they yield the classical Fresnel coefficients \cite{stratton}. 
\begin{equation}     \label{fresnel}
r^{s}_{a}=\frac{k_0-k_{a}}{k_0+k_{a}}  \ \     
r^{p}_{a}=\frac{\epsilon_{a}k_0-k_{a}}{\epsilon_{a}k_0+k_{a}},    
\end{equation}
 However, the surface impedances are more general, as they are also valid
for spatially dispersive systems, for which the Fresnel relations are not
applicable \cite{halevi,exciton}. 

In this paper we show that the use of  Eqs.(\ref{impeds}) and (\ref{impedp}) 
yields exact resuts for the Casimir forces using an expression for the force valid for arbitrary materials.  As a non-trivial application of the surface impedance approach we calculate the Casimir force between  two semi-infinite slabs with a non-local  dielectric response.  

 Consider two  slabs  $a=1,2$ parallel to the $xy$ plane within free space
and 
separated by a vacuum cavity $V$ of length $L$ along the $z$-direction, with
inner boundaries at 
$z_1=0$ and $z_2=L$ as shown in Fig.1. We assume that the slabs are
non-chiral, 
translational invariant and 
isotropic within the $xy$  
plane, but otherwise they 
may be arbitrary. A given photon within $V$ impinging upon a slab $a$ may be reflected
with a probability amplitude $r_a^\alpha$ which depends on its polarization $\alpha$,
acquiring a phase $kL$ as it moves on to the other slab. Otherwise, it may be transmitted into the material with a
probability $T_a^\alpha = 1-|r_a^\alpha|^2$ where it can be absorbed, exciting electronic or vibrational degrees of
freedom, or it can be transmitted into the vacuum beneath the slab, in any case, becoming lost forever (multiple
reflections within the slab are implictly accounted for in the reflection amplitudes $r_a^\alpha$. In thermodynamic
equilibrium there would be
photons coming from the outer vacuum and photons radiated by the materials themselves that woukd
compensate exactly for the photons from $v$ lost through absorption and transmission, appearing with a
probability $T^\alpha_a$ with no definite phase relation to the lost photons. Thus, the equilibrium radiation
within the cavity $V$ depends exclusively on its geometry, characterized by $L$, and on  reflection amplitudes
$r_a^\alpha$. Thus, we construct an auxiliary system $S^\prime$ made of two infinitesimal sheets 
at $z=z_a$, and we postulate that their reflection amplitudes are given exactly by the same 
amplitudes as those of the original slabs. This way, we assure that the radiation field within the
fictitious cavity $V^\prime$ corresponds to the real one. We further assume that in $S^\prime$ a photon
may be transmitted from $V^\prime$ into the vacuum outside with an amplitude $t^\alpha_a$. By choosing
$|r^\alpha_a|^2  + |t^\alpha_a|^2=1$ we make certain that energy is conserved without having to
account for any internal degrees of freedom within the fictitious sheets. 

We study first the case of $s$-polarized waves choosing $x-z$ as the plane of 
incidence. Then the incident electric field can be written as  
${\bf E}({\bf r},t)={\bf E_0} e^{i(Qx-\omega t)} \phi(z)$, with amplitude
${\bf E_0}=(0,E_y,0)$, and the magnetic field
${\bf B}= (B_x,0,B_z)$, is determined by Maxwell equations: $-i q B_x =  \partial_z E_y$,
and $ q B_z =  Q E_y$, with wavevector ${\bf q_\pm}=(Q,0,\pm k)$. The field component $E_y$ satisfies a one-dimensional
wave equation. The solution satisfying the boundary conditions at slab 1 interface ($z=0^+$) is
\begin{equation}\label{eym}
E_y^<(z)= e^{-ikz}+ r_1^s e^{ikz},
\end{equation}
while at slab 2 interface ($z=L^-$) we have
\begin{equation}\label{eyM}
E_y^>(z)= e^{ik(z-L)}+ r_2^s e^{-ik(z-L)},
\end{equation}
where the reflection amplitudes $r_a^\alpha$ are given by Eqs.( \ref{refl}). 

The electric Green's function can now be calculated as
\begin{equation}\label{ge}
G^E_{k^2}(z,z') = \frac{E_y^<(z_<) E_y^>(z_>)}{W} ,
\end{equation}
where $z_<$ and $z_>$ are the smaller and larger of $z$ and $z'$, respectively,
and $W$ is their Wronskian. 
 The 
magnetic Green's function is obtained by replacing $E_y\to B_x$ and
$r_a^s \to -r_a^s$ in
Eqs. (\ref{eyM})-(\ref{ge}). We do not consider $B_z$ separately, as
it is simply proportional to $E_y$. Therefore, for each ${\b Q}$, the local density of states 
per unit $k^2$ is given by \cite{plunien} 
\begin{equation}\label{rho2}
\rho^s_{k^2}(z) = -\frac{1}{2\pi} \mbox{Im} 
\left( G^E_{\tilde
k^2}(z,z)+ G^B_{\tilde k^2}(z,z)\right), 
\end{equation}
with $(\tilde k\equiv k+i0^+)$. 
By substituting Eqs. (\ref{ge})-(\ref{eyM}) and its magnetic analogues 
we obtain
\begin{equation}\label{rhos2}
\rho^s_{k^2}= \frac{1}{2 \pi \tilde k} \mbox{Re} \left[ \frac{1+r_1^s r_2^s e^{2i\tilde kL}}{1-r_1^s r_2^s e^{2i\tilde kL}} \right],
\end{equation}
which is independent of $z$. Given the symmetries of the problem, the $p$-polarization density of
states $\rho^p_{k^2}$ may be derived similarly, by just replacing $B_x \rightarrow -E_x$,
and $E_y \rightarrow B_y$. The final expression  is simply given by Eq.(\ref{rhos2}) after replacing all the superscripts $s\to p$. Finally, the total density of states is  
$\rho_{k^2} = \rho^s_{k^2}+\rho^p_{k^2}$.\\

Consider now a photon in a state characterized by $\alpha$, ${\bf Q}$, and $k^2$.
Its momentum is 
$\pm \hbar k$ and it moves with velocity $ \pm ck/q$ along the $z$ direction, 
so that its 
contribution to the momentum flux is $\hbar c k^2/q$. 
Denoting by $f(k)=N_k +1/2$ to photon occupation number of state $k$, 
the momentum flux from the vacuum gap into slab 2 is then given by
\begin{equation}
\sum_\alpha \int QdQ dk^2 \hbar c \frac{k^2}{q} f(k) \rho_{k^2}^\alpha. 
\end{equation}
There is a similar contribution coming 
from the semi-infinite vacuum on the other side of the slab, obtained by substituting 
$r_2^\alpha\to0$ above and reversing the flux direction $z\to-z$. 
The total force per unit area is obtained 
by subtracting the contributions from either side: 
\begin{equation}\label{F}
\frac{F(L)}{A}=\frac{\hbar c}{2 \pi ^2} \int_0^{\infty} dQ Q \int_{q\ge 0} dk 
\frac{\tilde k^3}{q} \mbox{Re} \frac{1}{k} \left[ 
\frac{1}{\xi^s-1)} +
\frac{1}{\xi^p-1} \right],
\end{equation}
with $\xi^\alpha = (r_1^\alpha r_2^\alpha exp(2ikL))^{-1}$.
The integral over $k$ runs from $iQ$ to 0 and then to $\infty$, so that $q$ remains real 
and positive. This expression depends only on the reflection coefficients or, 
equivalently, on the surface impedances of the system and the slab 
separation. Lifshitz formula is recovered upon   
substitution of the $local$ Fresnel amplitudes Eqs.(\ref{fresnel}),
whereas for perfect mirrors ($r_a^\alpha=\pm 1$) Eq.(\ref{F}) 
yields the expected Casimir force. 

An example of a non-trivial application of the impedance approach is the calculation of Casimir force between two conductors with a dielectric function that shows spatial dispersion. 
This is, $\epsilon({\vec r}-{\vec r',t-t'})$, or in the Fourier space $\epsilon({\bf q},\omega)$. 
Kats \cite{kats} studied nonlocal effects in an approximate way, and as he stated, it is necessary to specify correctly the dependence of the dielectric function on the wave vector. This is done in this section.   
In a conductor, the normal component of an incident  $p-polarized$ wave pushes the conduction charge towards or away of the surface creating an excess charge. When the frequency of the electromagnetic wave is greater than the plasma frequency $\omega_p$ of the metal, this charge accumulation propagates as a longitudinal wave (plasmon).  Thus, at a surface, $p-polarized$ waves couple to bulk plasmons taking away energy from the inicident wave and changing the values of the reflection amplitudes at the surface. 
 In this case the medium is said to be active and it supports both longitudinal and transverse oscillations.  We have to distinguish between the longitudinal dielectric function $\epsilon({\bf q},\omega)_l$ describing the response to the longitudinal dielectric field ${\bf E}_{||}$(parallel to the wave vector) and the transverse dielectric function $\epsilon({\bf q},\omega)_t$ that is the response to a transverse electric field. The total field is the sum of the longitudinal and transverse parts ${\bf E}={\bf E}_{||}+{\bf E}_{\bot}$.  Thus, in the metal we have a $p-polarized$ wave with wave vector $(Q,0,k)$ satisfying $k^2=\epsilon(\omega)_t \omega^2/c^2-k^2$ and a longitudinal wave with wavevector ${\bf l}=(Q,0,l)$ that obeys the relation $\epsilon({\bf l},\omega)_l=0$. Since $s-polarized$ waves do not couple to plasmons the transverse dielectric function is local (no spatial dispersion).
 To describe the dielectric response of the metal we have a Drude  transverse dielectric function $\epsilon_t=1-\omega_p^2/(\omega(\omega+i \gamma))$, where $\omega_p$ is the plasma frequency and $\gamma$ the damping.  For the longitudinal part we consider a hydrodynamic dielectric function \cite{cottam} given by
 \begin{equation}
 \epsilon^{l}(\vec{l},\omega)=1-\frac{\omega_p^2}{\omega^2+i \omega \gamma-\beta^2(Q^2+l^2)},
 \label{elong}
 \end{equation}
 where $\beta^2=3 v_f^2/5$ with $v_f$ the Fermi velocity of the metal \cite{cottam} and it enters the hydrodynamic model as a compressibility term of the charge carriers an it is responsible for the spatial dispersion \cite{barton}.  
 
 The surface impedance for $p-polarized$ waves can now be calculated at the surface of the metal. However, since we now have the additional longitudinal field,   we need an additional boundary condition.  For the problem we are considering this condition is that 
  the normal component of the total current be  zero at the surface.  This is,  $\vec{j}_{z\bot}+\vec{j}_{z ||}=0$.   The surface impedance is calculated as
  \begin{eqnarray}
  Z_p&=&\frac{E_{x\bot}+E_{x ||}}{B_y} \nonumber \\
         &=&\frac{kc}{\epsilon_t \omega}+\frac{(\epsilon_t-1)Q^2 c}{\epsilon_t l \omega}.
  \label{zpnolocal}
  \end{eqnarray}
 This expression for the impedance is now used in Eq.(\ref{refl}) to obtain the reflectivity for $p-polarized$ waves in the non-local case as
 \begin{equation}
 \label{reflp}
 r_p=\frac{\epsilon_{t}k_0-k-(\epsilon_t-1)Q^2/l}{\epsilon_{t}k_0+k+(\epsilon_t-1)Q^2/l},
 \end{equation}  
 while the reflectivity for the $s-polarized$ waves is the local Fresnel coefficient  $r_s$ given in Eq.(\ref{fresnel}). In the local limit ($Im(l)\rightarrow 0$) the local Fresnel coefficient is recovered. 
 
 The Casimir force between two semi-infinite conductors described by the HD model an now be calculated again using Eq.  (\ref{F}). In Figure 3 we present the percent difference between the Casimir force for the local case($F_{L}$) with that obtained with nonlocal effects ($F_{NL}$), i.e. $\Delta\%=|(F_{NL}-F_{L})/F_{L}|100$.  This is done for three different metals (K, Au, Al)  as a function of the separation of the slabs in units of the plasma wavelength of the metal $L/\lambda_p$. The parameters for  each metal ($\omega_p,\gamma, v_f$) are taken from the literature \cite{cocol,aschcroft}.   As expected the nonlocal effects become important for separations of the order of the plasma wavelength $\lambda_p$, and the difference between them can be significant for separations less than $\lambda_p$.  For these separations the vacuum modes that contribute mainly to the Casimir force have a frequency larger than $\omega_p$ thus having propagating modes in the metals. Furthermore the better the conductor the higher the difference between the local and nonlocal cases.  

The non-coincident results for the Casimir forces associated to the Lifshitz 
and the impedance approach Refs.\cite{kika} 
can be traced back to the introduction of
an approximate expression for the impedance. In the original work of Mostepanenko
and Trunov \cite{trunov}, they employed an expression for the 
impedance $Z(\omega) = 1/\sqrt{\epsilon(\omega)}$, valid only for small values 
of the electromagnetic wavevector ${\bf Q}$, parallel to the surface of incidence.
However, by introducing the definitions (\ref{impeds}) and (\ref{impedp})
in the same formalism, then the $exact$  expression for the Casimir force 
(Eq.(\ref{F})) is obtained 
by just replacing $Z \rightarrow Z^p$ in Eq.(41) of Ref.(\cite{kika}),  
and  $Z \rightarrow Z^s$ in Eq.(42) of the same reference.

In conclusion, we have shown that with the correct definitions of surface impedance an exact result for the Casimir force is obtained.  As case studies we calculated the surface impedance and Casimir force between an homogeneous slab with a non-local optical response.  In this case, we show that nonlocal effects become relevant for separations less than the plasma wavelength of the slab.  

Acknowledgement: Partial support for this work was provided by DGAPA-UNAM Proyect 116002.

 \begin{figure}[tbh]
\centerline {
}
\caption{Schematics of the system used in our calculations. For the nonlocal case, the slab has an additional longitudinal field as shown in the Figure.}
 \end{figure}

 \begin{figure}[tbh]
\centerline {
}
\caption{Percent difference between the Casimir force when local and nonlocal effects are considered. The separation between the slabs is in units of the plasma wavelength $\lambda_p$ of the metals. The curves correspond to Au, Al and K. The horizontal line is a visual aid to show where the percent difference is 1\%. }
 \end{figure}

\end{document}